\documentclass[11pt]{article}

\usepackage[a4paper,margin=1in]{geometry}
\usepackage{amsmath,amssymb,bm}
\usepackage{graphicx}
\usepackage{booktabs}
\usepackage{makecell}
\usepackage{float}
\usepackage{authblk}
\usepackage[numbers,sort&compress]{natbib}
\usepackage[colorlinks=true,linkcolor=blue,citecolor=blue,urlcolor=blue]{hyperref}

\graphicspath{{Figures/}}

\title{Design and Exploration of Ultra-Compact Quasi-Isodynamic Stellarator Configurations Using Domain-Adaptive Latent Diffusion Models}

\author[1,2]{Yang Han\textsuperscript{\textdagger}}
\author[1,2]{Hanlin Chen\textsuperscript{\textdagger}}
\author[1,2]{Shuai Cao}
\author[1]{Zhiyuan Lu}
\author[1,*]{Dehong Chen}
\author[1,*]{Guosheng Xu}
\author[1]{Baonian Wan}
\affil[1]{Institute of Plasma Physics, Chinese Academy of Sciences, Hefei 230031, China}
\affil[2]{University of Science and Technology of China, Hefei 230026, China}
\affil[*]{Email: dehong.chen@ipp.ac.cn, gsxu@ipp.ac.cn}

\date{}

\begin{document}

\maketitle
\begin{center}
\footnotesize\textsuperscript{\textdagger}These authors contributed equally to this work.
\end{center}

\begin{abstract}
Stellarator design explores a vast space of three-dimensional plasma boundaries, only a small fraction of which yields usable equilibria. Data-driven models can narrow this search by learning from existing optimized configurations. Building on the ConStellaration database, we extend conditional boundary generation to four-field-period QI configurations, focusing on the sparsely sampled low-aspect-ratio regime. The approach learns the common geometric structure of known QI equilibria and then adapts it using a small high-fidelity compact dataset, allowing target magnetic properties to guide generation beyond the original data distribution. This adaptation reduces the compact-domain test loss by approximately 87\% and yields converged ultra-compact candidates consistent with the prescribed conditions. Several candidates show favorable confinement indicators, and one provides a useful seed for further QI optimization and finite-\(\beta\) assessment. The method therefore serves as a data-informed front end to high-fidelity physics and optimization.
\end{abstract}


\section{Introduction}

Stellarators are magnetic-confinement fusion devices that confine high-temperature plasmas using
three-dimensional non-axisymmetric magnetic fields. Unlike tokamaks, whose poloidal confining field
is mainly produced by a strong toroidal plasma current, stellarators
rely primarily on externally shaped three-dimensional coils. They therefore offer
potential advantages for steady-state operation and for avoiding large-current-driven instabilities
\cite{boozer2021stellarators,beidler2021w7x}. The central design problem is to find three-dimensional plasma
boundaries that satisfy magnetohydrodynamic (MHD) equilibrium and engineering feasibility while
forming well-nested flux surfaces and providing favorable particle confinement. Quasisymmetry
(QS) and quasi-isodynamicity (QI) are two important optimization targets for
advanced stellarators. By controlling the structure of magnetic-field strength in
Boozer coordinates, they reduce bounce-averaged radial drifts of trapped particles
and neoclassical transport losses, thereby improving particle and energy confinement
\cite{helander2014theory,boozer1981plasma,nuhrenberg1988quasi,cary1997omnigenity,landreman2022precise,jorge2022single}.

Stellarator configuration design can usually be formulated as a high-dimensional
nonlinear optimization problem constrained by three-dimensional MHD equilibrium. Given boundary
Fourier coefficients, pressure profiles, rotational-transform profiles, and the number of
field periods, one must repeatedly call equilibrium solvers such as
VMEC and DESC to compute magnetic fields and equilibria, and
then evaluate aspect ratio, rotational transform, quasisymmetric or quasi-isodynamic residuals,
and other physical metrics, typically within optimization frameworks such as
SIMSOPT \cite{hirshman1983vmec,dudt2020desc,panici2023desc1,conlin2023desc2,landreman2021simsopt}. Because the boundary parameter space
is high-dimensional, the objective landscape is nonconvex, and the physical
constraints are strongly coupled, conventional optimization workflows require extensive equilibrium
solves and parameter scans. This cost makes large-scale exploration of
new QI configurations particularly challenging.

Data-informed representations offer a complementary strategy by constructing search
coordinates from databases of physically meaningful boundaries. For example, PCA and
quantile-based coordinates can improve the scaling and usable range of global stellarator
optimization \cite{landreman2026bayesianoptimizationstellaratoralphaparticle}. This highlights
the representation of the boundary space as a central design choice. The present work
focuses instead on learning a conditional distribution that can generate multiple
candidates from requested QI properties and adapt to a data-sparse target domain.

Recent generative machine-learning methods have demonstrated the ability to generate
candidate structures directly from target properties in molecular design, materials
design, and scientific inverse problems \cite{wang2023scientific,sanchez2018inverse}. Physics-informed neural networks have
also begun to improve the efficiency and accuracy of MHD-equilibrium-related
calculations \cite{thun2026pinn}. Introducing such methods into stellarator design offers a
complementary route: learning statistical maps between physical indicators and three-dimensional
boundary shapes, rapidly generating candidates from target magnetic-geometric conditions, and
shifting expensive MHD equilibrium calculations from iterative search to post-generation
validation and screening.

Existing studies have provided important foundations for data-driven stellarator design.
Ref.~\cite{padidar2025diffusion} trained diffusion models on the QUASR quasisymmetric stellarator dataset
for quasi-axisymmetric (QA) and quasi-helically symmetric (QH) configurations, demonstrating the
potential of generative models for reproducing target magnetic-geometric indicators and
quasisymmetric properties. Ref.~\cite{cadena2025constellaration} released ConStellaration, a dataset of QI stellarator
plasma boundaries, MHD equilibria, and related performance metrics, providing an
open benchmark for data-driven QI research. Ref.~\cite{curvo2025deep} used mixture density
networks to generate high-aspect-ratio stellarator configurations, further showing that probabilistic
generative models can support inverse design in high-dimensional boundary spaces. Together,
these studies motivate treating the empirical distribution of usable boundaries as a
prior to be represented, conditioned, and subsequently tested by high-fidelity physics.

Compared with QS configurations, QI configurations belong to the broader
class of omnigenous magnetic-field designs. Under ideal omnigenity, the bounce-averaged
radial drift of trapped particles vanishes. QI configurations can be
viewed as a class of realizations in which contours of
magnetic-field strength in Boozer coordinates are approximately poloidally closed, and
they often possess low neoclassical transport and favorable energetic-particle confinement
potential \cite{cary1997omnigenity,landreman2012omnigenity,goodman2023constructing,goodman2024prx,liu2025omnigenity,rodriguez2026database}. The optimization experience of W7-X and recent reactor-relevant
QI design studies indicate that QI design principles are important
for improving stellarator confinement performance \cite{beidler2021w7x,jorge2022single,goodman2024prx,goodman2025reactor,lion2025stellaris,liu2026low,sanchez2026ciemqa}. However, compared with existing
generative-model studies for QA/QH configurations, conditional generation and domain generalization
for QI configurations, especially ultra-compact QI configurations, remain insufficiently explored.

This work focuses on four-field-period QI stellarator configurations and proposes
a rapid generation framework based on variational autoencoder (VAE) latent
representations and conditional diffusion models. We further combine this framework
with domain adaptation to explore ultra-compact low-aspect-ratio design regimes. The
main contributions are as follows:
\begin{itemize}
    \item We construct a VAE latent representation with a physical-space \(R/Z\)
          reconstruction constraint. This representation significantly compresses high-dimensional boundary Fourier coefficients
          while preserving the true three-dimensional boundary geometry, providing a compact
          and physically consistent low-dimensional representation for conditional QI generation.
    \item Building on existing generative inverse-design ideas for stellarators, we extend
          conditional diffusion generation to latent-space scanning of ultra-compact QI configurations.
          The resulting model can rapidly produce physically verifiable candidate boundaries
          for low-aspect-ratio target regions and thereby support design-space exploration in
          early-stage stellarator optimization.
    \item We propose a domain-adaptation strategy for the high-fidelity, small-sample ultra-compact
          QI regime. By combining full-model low-learning-rate fine-tuning with source-domain replay,
          the generative prior learned from large-scale ordinary QI data is
          transferred to low-aspect-ratio and low-rotational-transform regions insufficiently covered by the
          original HL set, a high-fidelity optimized-configuration database, thereby providing more focused candidate initial conditions
          for subsequent high-fidelity QI optimization.
    \item Through two-dimensional parameter scans, DESC equilibrium verification, effective-ripple evaluation, and
          particle-loss diagnostics, we identify ultra-compact candidate configurations with strong confinement
          potential. In particular, some configurations obtained lie outside the compactness range covered
          by the original HL set while maintaining nearly zero core
          \(\alpha\)-particle loss, demonstrating the ability of the framework to discover
          previously unexplored candidates that can serve as initial conditions for
          subsequent high-\(\beta\) optimization.
\end{itemize}

\section{Data and Methods}

The QI stellarator configuration data used in this work consist
of two parts. The first part is derived from the
ConStellaration open dataset, denoted here as the C set. We
select samples with toroidal field period \(N_{\mathrm{fp}}=4\) as the main
training source, increase the boundary Fourier resolution from \((M,N)=(4,4)\),
used in the original ConStellaration data, to \((M,N)=(6,6)\), recompute
fixed-boundary DESC equilibria at this higher resolution, and remove anomalous samples,
resulting in approximately \(2.7\times10^4\) physically consistent
candidate configurations \cite{cadena2025constellaration,hirshman1983vmec,dudt2020desc,panici2023desc1,conlin2023desc2}.
The second part is the original HL set, a high-fidelity QI optimized-configuration
database containing 409 configurations. This original HL set was
independently optimized in this work and serves as the physical reference
for the target domain. The HL
augmented set, constructed near the original HL configurations using DESC
equilibrium perturbations, is used only for domain-adaptation training, validation, and
noise-prediction-error evaluation. Generated configurations are evaluated in terms of equilibrium
convergence, target magnetic-geometric indicator errors, ultra-compactness, and Boozer-plot properties.

\subsection{Data Preprocessing}

Under stellarator symmetry, each plasma boundary is represented by the
following double Fourier series:
\begin{align}
R_b(\theta,\phi) &= \sum_{m,n} R_{mn}\cos\left(m\theta-nN_{\mathrm{fp}}\phi\right), \\
Z_b(\theta,\phi) &= \sum_{m,n} Z_{mn}\sin\left(m\theta-nN_{\mathrm{fp}}\phi\right),
\end{align}
where \(\theta\) and \(\phi\) are the poloidal and geometric toroidal angles,
respectively, and \(N_{\mathrm{fp}}=4\) is the number of field periods. The
parity of the cosine and sine terms enforces stellarator symmetry. The Fourier expansion is
truncated at \((M,N)=(6,6)\). After preprocessing, each sample contains 169 boundary Fourier
coefficients and four magnetic-geometric indicators, giving a data dimension of
\((27610,173)\). The four conditioning indicators are aspect ratio, average triangularity,
maximum elongation, and edge rotational transform. Their distributions are shown
in Fig.~\ref{fig:dataset_distribution}.

\subsection{Data Augmentation}

\begin{figure}[t]
    \centering
    \includegraphics[width=0.70\textwidth]{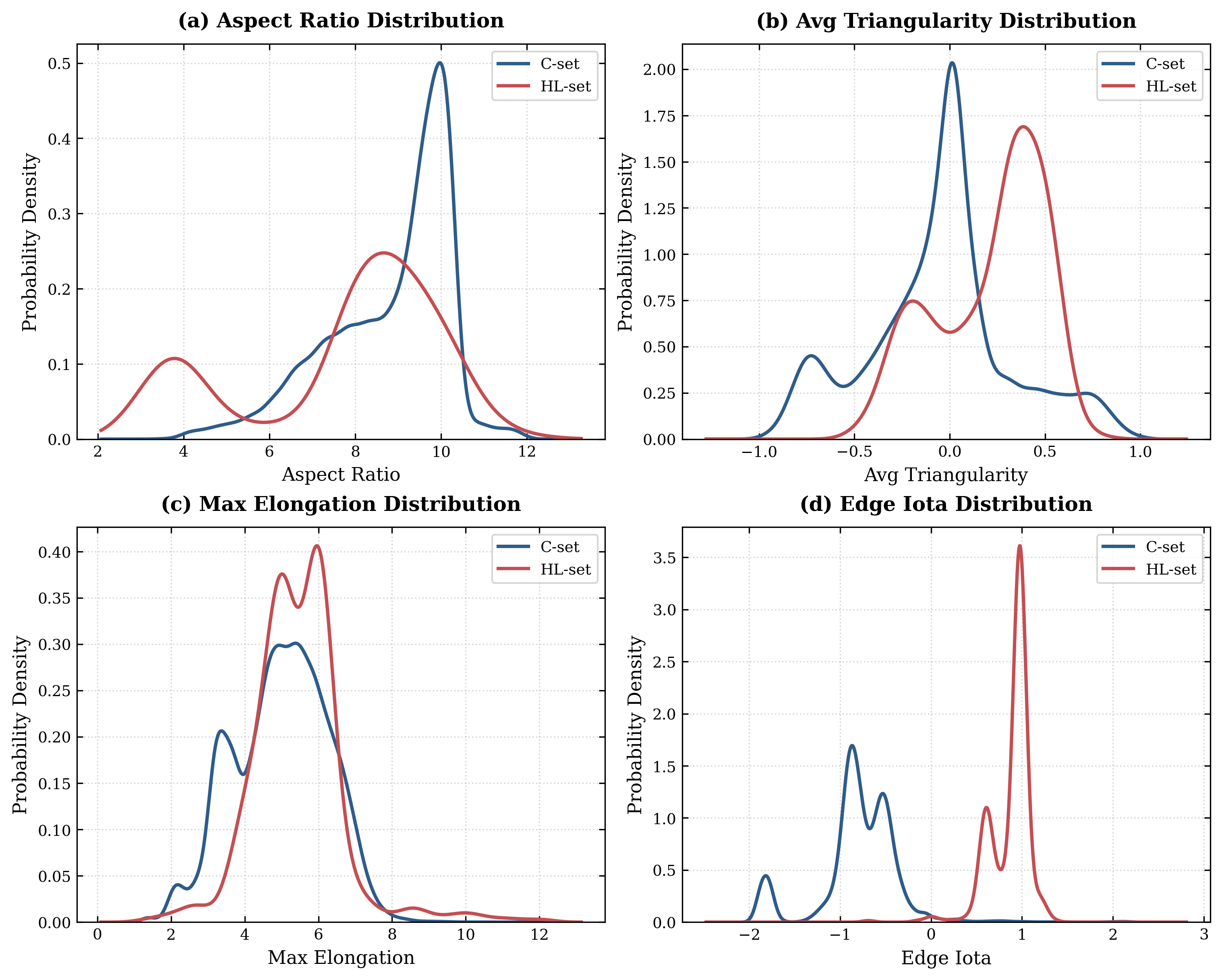}
    \caption{Kernel density estimation (KDE) distributions of magnetic-geometric indicators in the preprocessed C set and the original HL set.}
    \label{fig:dataset_distribution}
\end{figure}

To mitigate the limited size of the original HL set,
we use the DESC equilibrium perturbation method to perform physically
consistent data augmentation around converged fixed-boundary MHD equilibria \cite{conlin2023desc2}. The
method applies small perturbations to the original equilibrium and its
control parameters, and uses the local response of the equilibrium
equation to approximate the perturbed equilibrium. In practice, small relative
perturbations are applied to boundary Fourier coefficients, pressure profiles, rotational-transform
profiles, and total toroidal flux, followed by a DESC solve
to restore force balance. The augmented HL set contains \((1619,173)\)
samples and is used only for domain-adaptation-related training and error
evaluation. Figure~\ref{fig:dataset_distribution} shows the distributional differences between the C set
and the original HL set in the four magnetic-geometric indicators.

\subsection{Low-Dimensional Latent Compression of Boundary Coefficients}

To map high-dimensional boundary parameters to a low-dimensional continuous representation
suitable for diffusion modeling, we use a VAE to nonlinearly
compress the boundary Fourier coefficients \cite{kingma2014vae}. For each configuration, the
input is \(\bm{x}\in\mathbb{R}^{169}\), where the first 85 dimensions are \(R\)-direction
Fourier coefficients and the remaining 84 dimensions are \(Z\)-direction Fourier
coefficients. The VAE encoder maps the standardized boundary coefficients \(\tilde{\bm{x}}\)
to a low-dimensional latent space and defines the approximate posterior
\begin{equation}
q_{\phi}(\bm{z}|\tilde{\bm{x}})=
\mathcal{N}\left(\bm{\mu}_{\phi}(\tilde{\bm{x}}),
\mathrm{diag}\left(\bm{\sigma}_{\phi}^{2}(\tilde{\bm{x}})\right)\right).
\end{equation}
The decoder reconstructs the standardized boundary coefficients from the latent
variable. In the subsequent diffusion model, we use the posterior
mean \(\bm{\mu}_{\phi}(\tilde{\bm{x}})\) as a deterministic latent representation. To avoid boundary-shape
distortion caused by optimizing only in coefficient space, we add
a physical-space \(R/Z\) collocation-point reconstruction error to the VAE loss:
\begin{equation}
\mathcal{L}_{RZ}
=\left\langle (\hat R-R)^2+(\hat Z-Z)^2 \right\rangle_{\theta,\phi},
\end{equation}
where \(\langle\cdot\rangle_{\theta,\phi}\) denotes the average over boundary collocation points. The
final training objective is
\begin{equation}
\mathcal{L}
=\left\|\hat{\tilde{\bm{x}}}-\tilde{\bm{x}}\right\|_2^2
+\lambda_{RZ}\mathcal{L}_{RZ}
+\beta_{KL}D_{KL}\left(q_{\phi}(\bm{z}|\tilde{\bm{x}})\|\mathcal{N}(0,I)\right).
\end{equation}
Training is performed in two stages: the model first learns
the main statistical structure of the standardized Fourier coefficients, and
the \(R/Z\) reconstruction-loss weight is then gradually increased to balance
coefficient reconstruction accuracy and true boundary-geometry fidelity. To select a
latent dimension suitable for subsequent diffusion modeling, we perform a
representative scan over \(d_z=8,16,24,32,43,64\), and compare VAE performance with principal
component analysis (PCA) at the same dimensions. The evaluation metrics
include standardized Fourier-coefficient mean squared error (MSE) and physical boundary-space
\(R/Z\) root mean squared error (RMSE). Here PCA is used only as a linear
reconstruction baseline at the same latent dimensions.

\subsection{Conditional Diffusion Model in the VAE Latent Space}

After obtaining the low-dimensional latent representation, we construct a conditional
denoising diffusion probabilistic model (CDDPM) to learn the generative mapping
from target magnetic-geometric conditions to boundary latent variables \cite{ho2020ddpm,rombach2022latent}. A
training sample is written as \((\bm{z}_0,\bm{c})\), where \(\bm{z}_0\in\mathbb{R}^{43}\) is the
VAE-encoded boundary latent variable and \(\bm{c}\) contains aspect ratio, average
triangularity, maximum elongation, and edge rotational transform \(\iota\). Both latent
and condition variables are standardized using training-set statistics.

The forward diffusion process gradually adds Gaussian noise in latent
space. Given time step \(t\) and noise \(\bm{\epsilon}\sim\mathcal{N}(0,I)\), the noisy
latent variable is
\begin{equation}
\bm{z}_t=\sqrt{\bar\alpha_t}\bm{z}_0+\sqrt{1-\bar\alpha_t}\bm{\epsilon},
\end{equation}
where \(\alpha_t=1-\beta_t\), \(\bar\alpha_t=\prod_{s=1}^{t}\alpha_s\), and the noise variance \(\beta_t\) follows a
linear schedule. The reverse denoising network \(\epsilon_\theta(\bm{z}_t,t,\bm{c})\) is a conditional
multilayer perceptron (MLP). It takes the noisy latent variable, time-step
embedding, and condition vector as inputs, and outputs a noise
prediction with the same dimension as the latent variable.

\begin{figure}[t]
    \centering
    \includegraphics[width=\textwidth]{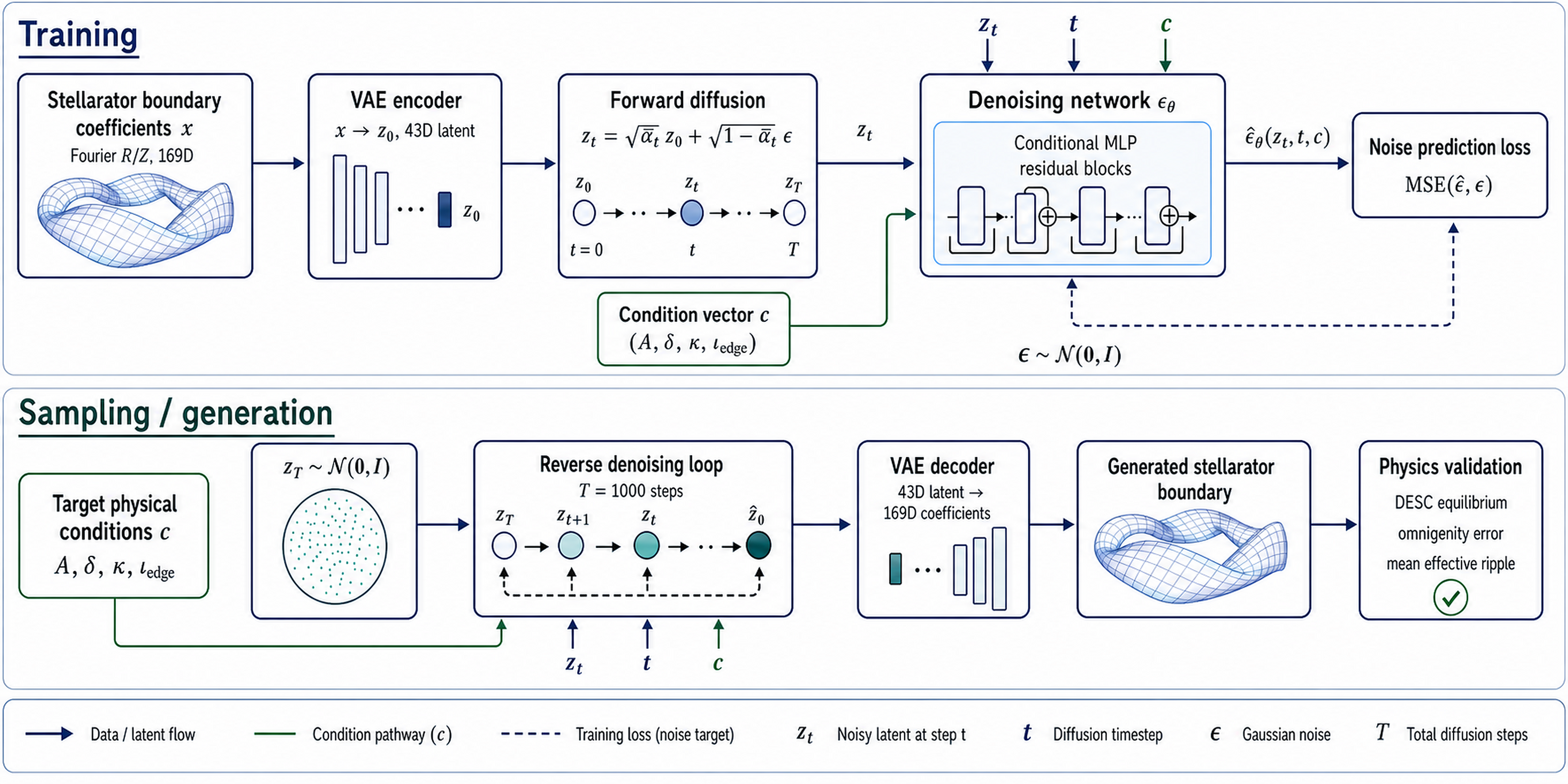}
    \caption{Architecture of the conditional denoising diffusion probabilistic model (CDDPM) in the VAE latent space.}
    \label{fig:cddpm_architecture}
\end{figure}

During training, \(\bm{z}_0\) and \(\bm{c}\) are sampled from the training
set, while \(t\) and \(\bm{\epsilon}\) are randomly sampled. The model
is optimized by minimizing the MSE between predicted and true
noise:
\begin{equation}
\mathcal{L}_{\mathrm{CDDPM}}=
\mathbb{E}_{\bm{z}_0,\bm{c},t,\bm{\epsilon}}
\left[
\left\|\epsilon_\theta(\bm{z}_t,t,\bm{c})-\bm{\epsilon}\right\|_2^2
\right].
\end{equation}
The model is trained with AdamW, and the best weights
are selected according to validation noise-prediction error. During sampling, generation
starts from standard Gaussian noise. Under a given target condition
\(\bm{c}\), the model performs reverse denoising to obtain \(\hat{\bm{z}}_0\), which
is decoded by the VAE decoder into three-dimensional boundary Fourier
coefficients. The CDDPM architecture is shown in Fig.~\ref{fig:cddpm_architecture}.

\subsection{Domain Adaptation for High-Fidelity and Ultra-Compact QI Configurations}

More compact, lower-aspect-ratio stellarators are attractive from both engineering and
physics perspectives. Reducing the aspect ratio can decrease the major
radius, magnet size, and radial build required for a given
fusion power, thereby improving device economy and accessibility. At the
same time, stronger toroidal shaping in compact configurations can modify
curvature and local magnetic shear, providing a potential route to
improved ballooning-mode stability at high pressure \cite{najmabadi2006compact,najmabadi2008aries,najmabadi2008iaea,helander2014theory}. QI configurations,
which may also be viewed as poloidally omnigenous configurations, are
particularly appealing because they can combine low neoclassical transport, favorable
fast-particle confinement, and stellarator operation with nearly zero net toroidal
plasma current. This current-free character strongly reduces susceptibility to large-current-driven
disruptive events while preserving the steady-state advantages of stellarators \cite{goodman2024prx,goodman2025reactor,lion2025stellaris,liu2026low,sanchez2026ciemqa}.
Thus, the ultra-compact QI design space is attractive from both
physics and engineering perspectives.

More importantly, this attractive design space often lies near or outside
the boundary of existing high-fidelity sample distributions. Traditional point-wise optimization
has difficulty systematically covering low-aspect-ratio and low-rotational-transform combinations, especially when
compactness, equilibrium convergence, and confinement-related constraints must be satisfied simultaneously.
If a generative model can produce converged and target-consistent candidate
boundaries in regions insufficiently covered by the HL set, it
can provide new initial configurations for subsequent high-fidelity QI optimization. Because
the high-fidelity ultra-compact QI dataset differs substantially from the original
training set, directly applying a conditional diffusion model trained on
ordinary-aspect-ratio samples cannot accurately cover the low-aspect-ratio design region. We
therefore use a fine-tuning-based domain-adaptation strategy \cite{pan2010transfer,yosinski2014transferable} to transfer the
pretrained boundary-generation capability to the target domain.

We compare three fine-tuning strategies. The first freezes the main
denoising network and updates only the condition embedding layer and
final output layer, testing whether the model can adapt to
the new domain by adjusting the condition mapping while preserving
the original denoising structure. The second performs full-model fine-tuning with
a small learning rate, allowing all parameters to adapt to
the new domain while reducing the risk of catastrophic forgetting
\cite{kirkpatrick2017forgetting}. The third combines full-model low-learning-rate fine-tuning with source-domain replay,
mixing a fraction of original-domain samples into the new-domain training
set to constrain the model from losing its source-domain generative
capability \cite{shin2017replay}. All strategies are initialized from the same pretrained
conditional diffusion model and use AdamW, gradient clipping, and cosine
learning-rate annealing. Model selection is based on validation loss, and
domain-adaptation performance is evaluated by comparing target-domain test losses before
and after fine-tuning.

\subsection{Downstream Vacuum QI Optimization and Finite-\texorpdfstring{\(\beta\)}{beta} Assessment}

To assess the downstream utility of a generated ultra-compact vacuum
configuration, we use \(\mathrm{eq\_0046}\) as the initial condition for a
vacuum QI boundary optimization in DESC. The objective combines current-density,
omnigenity-harmonic, magnetic-well, and OOPS-mapping-overlap terms. All boundary
Fourier coefficients are free except the \((m,n)=(0,0)\) radial coefficient,
whereas the pressure profile, current profile, and toroidal flux are fixed. No
aspect-ratio objective or constraint is imposed in this 200-iteration vacuum
optimization.

The optimized vacuum boundary is subsequently used for a fixed-boundary
finite-\(\beta\) equilibrium calculation. A prescribed pressure profile is
applied, and the boundary, pressure profile, zero net toroidal current, and
toroidal flux are held fixed while the MHD force-balance residual is minimized.
The resulting finite-\(\beta\) equilibrium is assessed using Boozer-coordinate
magnetic-field contours, NEO effective ripple, SIMPLE \(\alpha\)-particle
tracing, and COBRA ideal-ballooning calculations. The SIMPLE calculations use
500 particles at \(s=0.01\) and \(s=0.25\), tracked to \(0.2~\mathrm{s}\)
after scaling the minor radius and magnetic field to \(1.7~\mathrm{m}\) and
\(5.7~\mathrm{T}\), respectively. COBRA evaluates the maximum ideal-ballooning
eigenvalue over discrete radial, poloidal, and field-line samples.

\section{Results}

\subsection{VAE Dimensionality Selection and Boundary Reconstruction}

\begin{figure}[t]
    \centering
    \includegraphics[width=\textwidth]{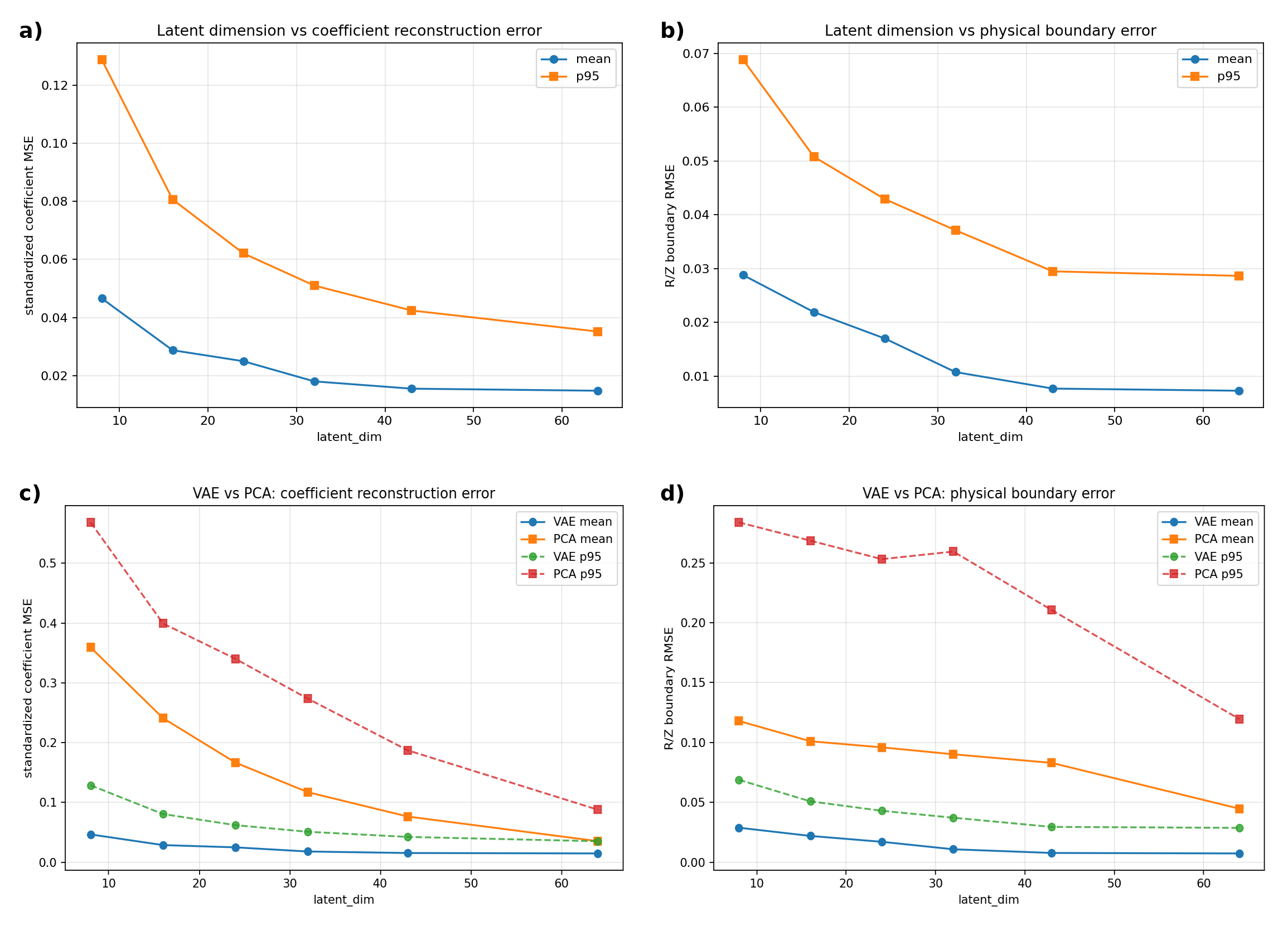}
    \caption{Comparison of VAE and PCA boundary reconstruction errors across different reduced dimensions. (a) VAE coefficient-space MSE; (b) VAE physical-space \(R/Z\) boundary RMSE; (c) coefficient-space MSE comparison between VAE and PCA; (d) physical-space \(R/Z\) boundary RMSE comparison between VAE and PCA.}
    \label{fig:vae_pca}
\end{figure}

We first evaluate the nonlinear dimensionality-reduction error of the VAE
at different latent dimensions and compare it with PCA at
the same dimensions, as shown in Fig.~\ref{fig:vae_pca}. Figures~\ref{fig:vae_pca}c and \ref{fig:vae_pca}d
show that VAE reconstruction errors are significantly lower than PCA
errors across all latent dimensions, both in terms of average
error and the 95th percentile error that characterizes the high-error
tail. This indicates that the nonlinear VAE latent representation more
effectively preserves the main geometric features of stellarator boundaries. Figures~\ref{fig:vae_pca}a
and \ref{fig:vae_pca}b further show that both coefficient-space error and physical-space
\(R/Z\) boundary error decrease as the latent dimension increases, but
the improvement becomes more gradual at higher dimensions. We select
several representative candidate dimensions rather than performing an exhaustive uniform
grid search. Considering compression ratio, reconstruction accuracy, and the complexity
of the subsequent diffusion model, we use a 43-dimensional latent
space for the generation model, which provides a reasonable balance
between boundary fidelity and model complexity.

\begin{figure}[t]
    \centering
    \includegraphics[width=0.70\textwidth]{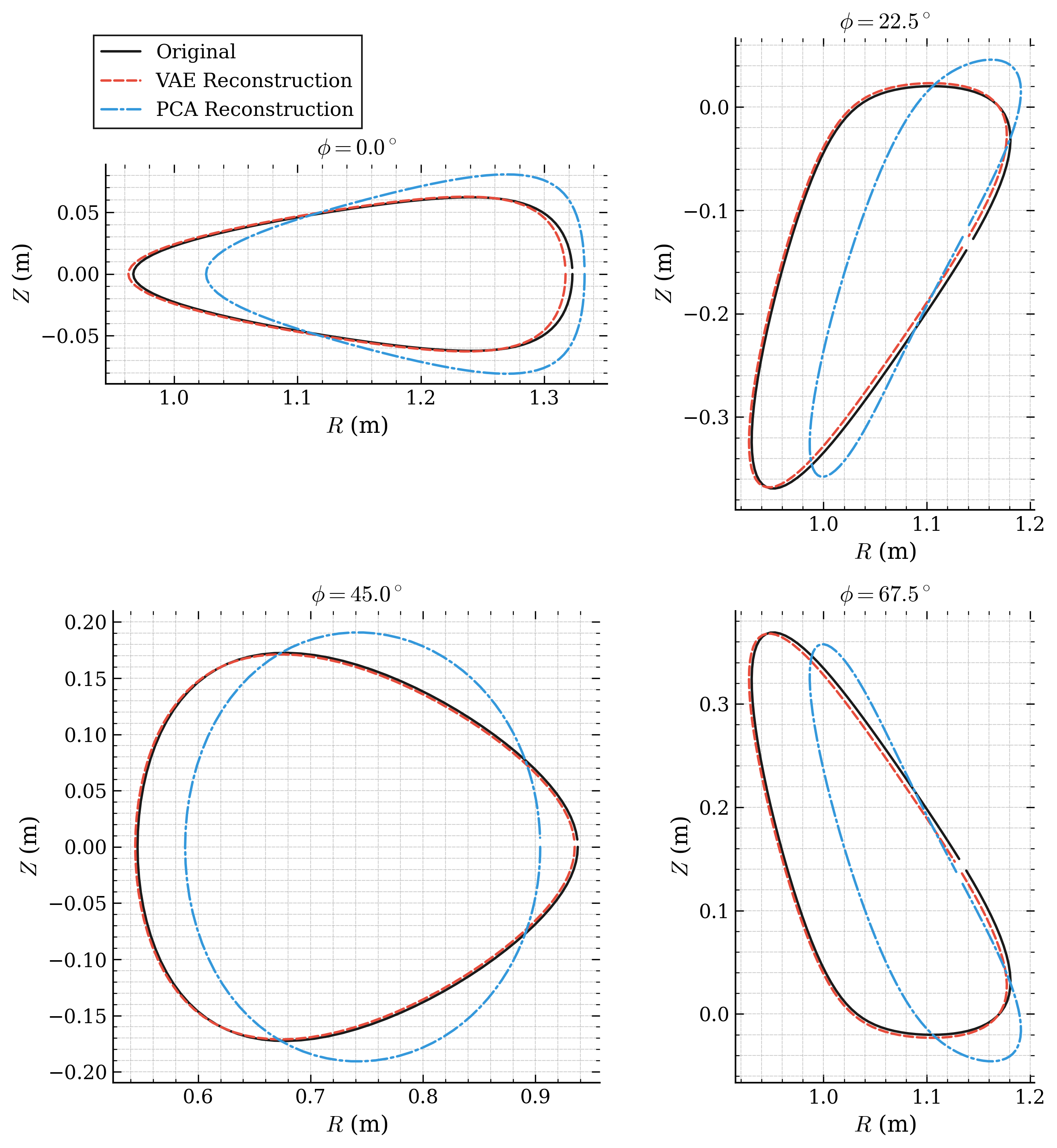}
    \caption{Boundary reconstruction comparison after reducing VAE and PCA representations to 43 dimensions.}
    \label{fig:boundary_reconstruction}
\end{figure}

To further examine actual boundary reconstruction at the 43-dimensional latent-space
working point, we select a representative test configuration and compare three-dimensional
boundary reconstructions by VAE and PCA in Fig.~\ref{fig:boundary_reconstruction}.
Quantitative evaluation in the physical $(R,Z)$ space shows that the VAE reconstructs the
boundary with a normalized error of approximately $0.33\%$ relative to the major radius
($R_{\text{maj}}\approx1.0$~m), i.e. a mean geometric deviation of approximately $3$~mm and
a maximum deviation of approximately $6$~mm. By comparison, PCA gives approximately
$4.1\%$, corresponding to mean and maximum deviations of approximately $4$~cm and
$8$~cm, respectively. These results confirm that the VAE achieves sufficient geometric
accuracy for subsequent conditional diffusion modeling, while outperforming PCA in both
average error and representative boundary visualization.

\subsection{VAE-Diffusion Training and Prediction Results}

\begin{figure}[t]
    \centering
    \includegraphics[width=0.80\textwidth]{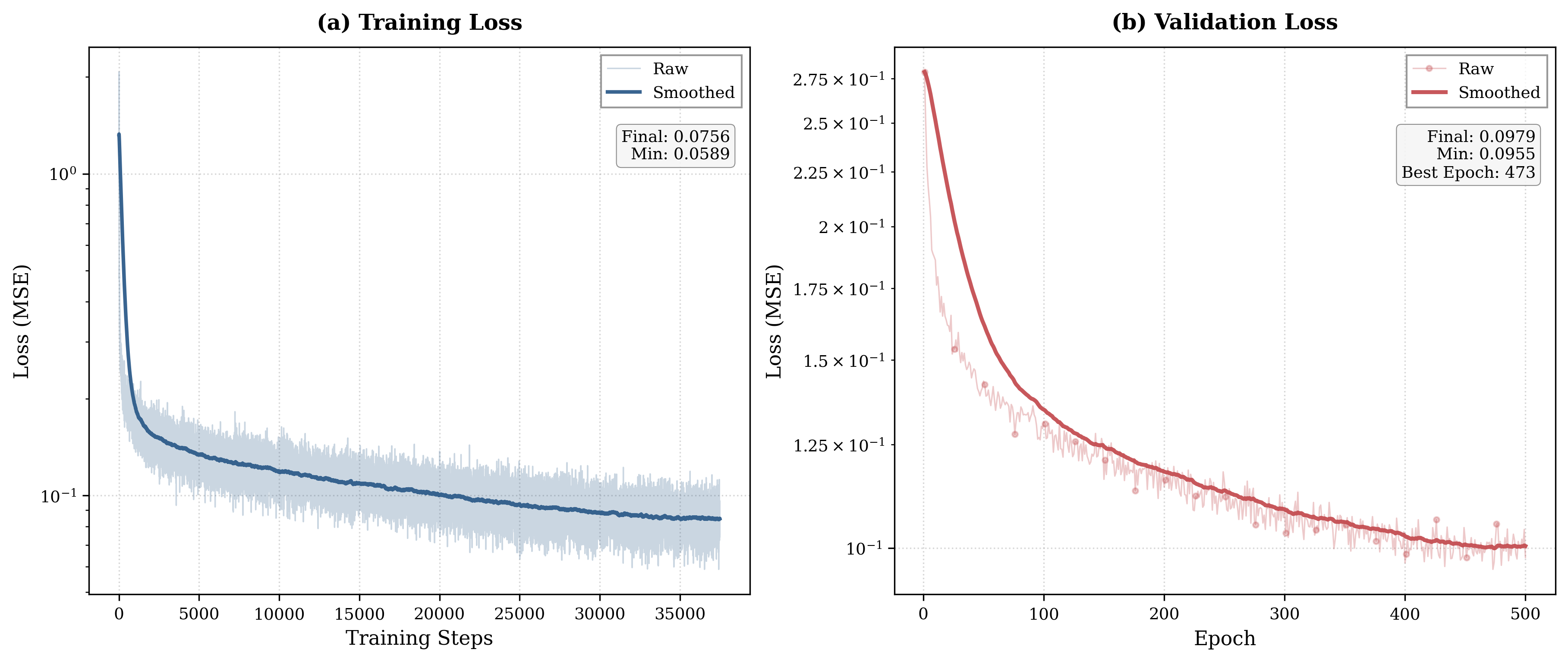}
    \caption{Noise-prediction MSE during CDDPM training and validation. Transparent thin lines show raw losses; darker thick lines show exponential moving average (EMA) smoothed curves.}
    \label{fig:training_curves}
\end{figure}

As shown in Fig.~\ref{fig:training_curves}, the conditional diffusion model exhibits stable
training convergence. The training-set noise-prediction MSE decreases rapidly at the
beginning and then continues to decrease slowly, reaching a final
training loss of approximately 0.0756 and a minimum training loss
of approximately 0.0589. The validation loss also drops quickly at
early epochs and then becomes stable, with a final validation
loss of approximately 0.0979 and a minimum validation loss of
approximately 0.0955 at about epoch 473. Together with the test-set
noise-prediction error of approximately 0.1004, the close validation and test
error levels indicate consistent performance on unseen samples and no
obvious overfitting. These results show that CDDPM can effectively learn
the conditional denoising map from target magnetic-geometric conditions to boundary
latent variables.

\begin{figure}[t]
    \centering
    \includegraphics[width=0.80\textwidth]{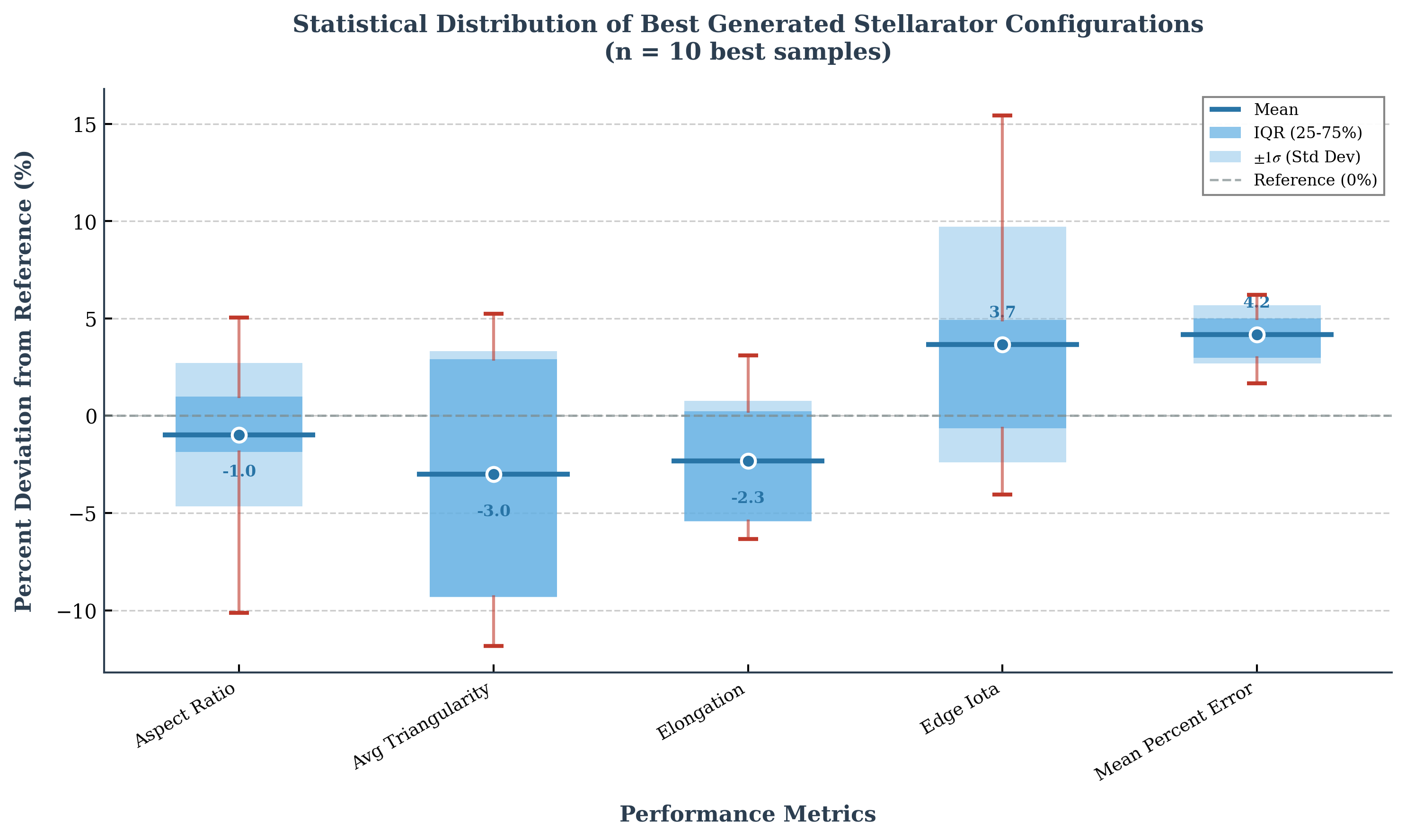}
    \caption{Error distributions of generated samples under prescribed magnetic-geometric target conditions.}
    \label{fig:condition_errors}
\end{figure}

We then use magnetic-geometric indicators from unseen test samples as
target conditions, generate candidate boundaries using the trained CDDPM, and
compute their realized magnetic-geometric indicators after DESC fixed-boundary equilibrium solves.
Figure~\ref{fig:condition_errors} shows the percentage deviations of generated samples relative to
the target conditions. The average deviations in aspect ratio, average
triangularity, maximum elongation, and edge rotational transform are approximately \(-1.0\%\),
\(-3.0\%\), \(-2.3\%\), and \(3.7\%\), respectively, all close to the zero-reference
line. The combined mean percentage error over the four indicators
is approximately \(4.2\%\). This indicates that the model maintains good
conditional consistency on unseen test conditions and can generate candidate
QI configurations that approximately match target magnetic-geometric parameters.

\subsection{Domain Adaptation and Equilibrium Assessment for Ultra-Compact QI Configurations}

\begin{figure}[t]
    \centering
    \includegraphics[width=0.80\textwidth]{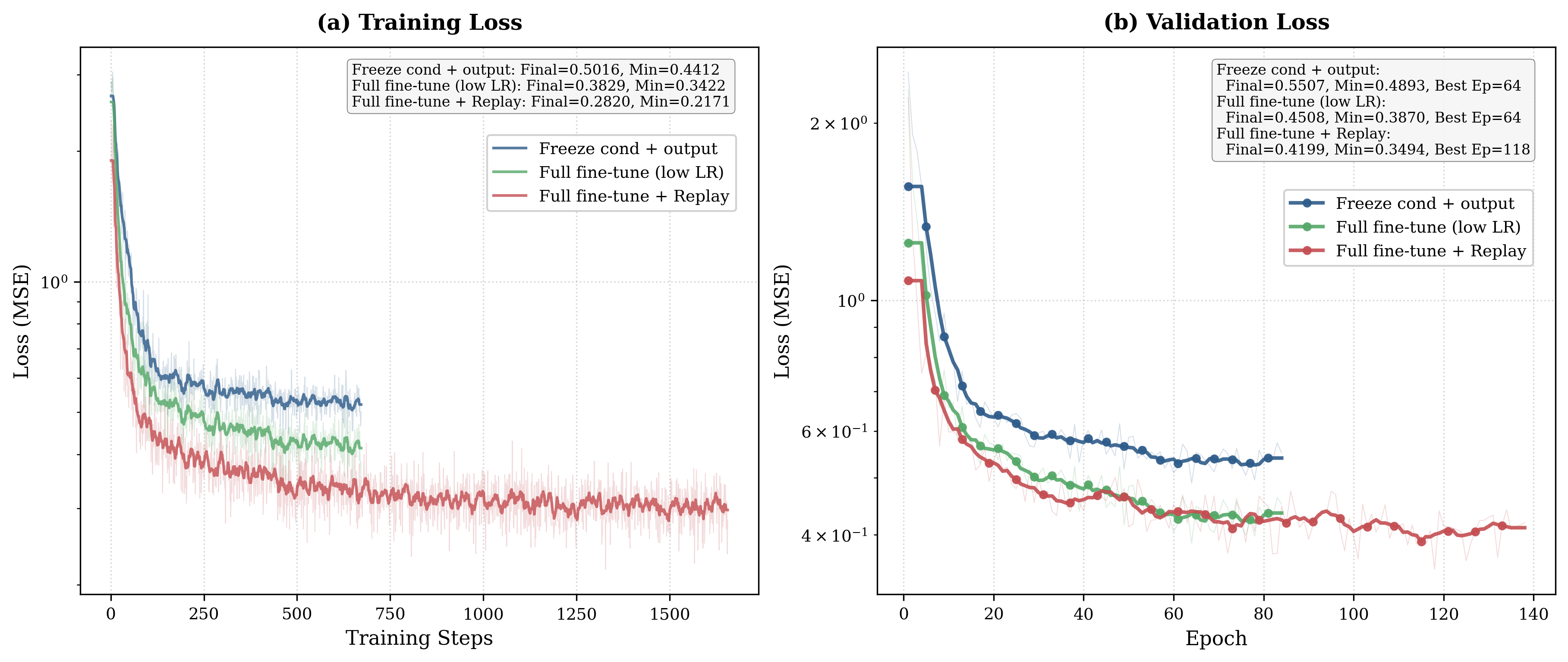}
    \caption{Domain-adaptation training results for the three fine-tuning strategies. (a) Training MSE as a function of training steps. (b) Validation MSE as a function of epoch. In both panels, the blue, green, and red curves correspond, respectively, to frozen-backbone fine-tuning that updates only the condition embedding and output mapping layers, full-model fine-tuning with a low learning rate, and full-model low-learning-rate fine-tuning with 30\% source-domain replay.}
    \label{fig:domain_training}
\end{figure}

To mitigate the distribution shift of the pretrained conditional diffusion
model in ultra-compact QI configurations with aspect ratio \(A<5\),
we compare three domain-adaptation strategies: freezing the model backbone and
updating only the condition embedding and output mapping layers; full-model
end-to-end fine-tuning with a small learning rate; and full-model low-learning-rate
fine-tuning with 30\% source-domain replay samples. The training and validation
loss curves are shown in Fig.~\ref{fig:domain_training}. All three strategies rapidly
reduce training loss within the first tens of epochs and
then gradually converge. The first two strategies converge at similar
speeds, while the replay strategy trains for more epochs and
exhibits smoother validation loss decay, ultimately reaching the lowest validation
loss.

Table~\ref{tab:domain_adaptation} summarizes the best validation loss, best epoch, and target-domain
test loss for the three strategies. The pretrained model has
a test loss of approximately 3.2596 on the compact target
test set. After fine-tuning, the test losses decrease to approximately
0.5644, 0.4450, and 0.4147 for the three strategies, corresponding to
relative improvements of 82.69\%, 86.35\%, and 87.28\%, respectively. These results
show that updating only the condition and output layers already
provides substantial improvement; full-model low-learning-rate fine-tuning further reduces validation and
test errors; and source-domain replay gives the best adaptation while
retaining source-domain knowledge. Although exact numerical values may vary slightly
due to training/validation splits and replay sampling randomness, the relative
ranking of the three strategies is stable across runs.

\begin{table}[t]
    \centering
    \small
    \caption{Fine-tuning performance comparison of three domain-adaptation strategies.}
    \label{tab:domain_adaptation}
    \vspace{2pt}
    \begin{tabular}{lccccc}
        \toprule
        Strategy & \makecell{Best\\epoch} & \makecell{Best val.\\loss} & \makecell{Pre-FT\\test loss} & \makecell{Post-FT\\test loss} & Improvement \\
        \midrule
        Freeze cond. + output & 63  & 0.489257 & 3.259626 & 0.564367 & 82.69\% \\
        Full fine-tune (low LR) & 63 & 0.387013 & 3.259626 & 0.444991 & 86.35\% \\
        Full fine-tune + replay & 117 & 0.349406 & 3.259626 & 0.414655 & 87.28\% \\
        \bottomrule
    \end{tabular}
\end{table}

To evaluate the adapted model's generative capability in the ultra-compact
region, we perform a two-dimensional scan in the parameter plane
spanned by aspect ratio \(A\) and edge rotational transform \(\iota\).
The aspect ratio characterizes device compactness, while \(\iota\) reflects magnetic
topology and rational-surface distribution. Together, they capture important trade-offs among
geometric scale, transport properties, and engineering feasibility in ultra-compact configuration
design.

For this scan and the subsequent representative-configuration screening, we
evaluate the omnigenity error using the Optimized Parallels (OOPS) formulation
of Liu \emph{et al.}\ \cite{liu2025omnigenity}, as implemented in DESC. The
reported quantity is the mean absolute residual in OOPS coordinates; the
coordinate mapping and complete mathematical definition are given in
Appendix~\ref{app:omni_error}. Because this quantity is evaluated within a
single optimization workflow, it is used only as an auxiliary relative
diagnostic and not to establish an absolute QI-quality ranking across
independently optimized configurations.

\begin{figure}[t]
    \centering
    \includegraphics[width=0.95\textwidth]{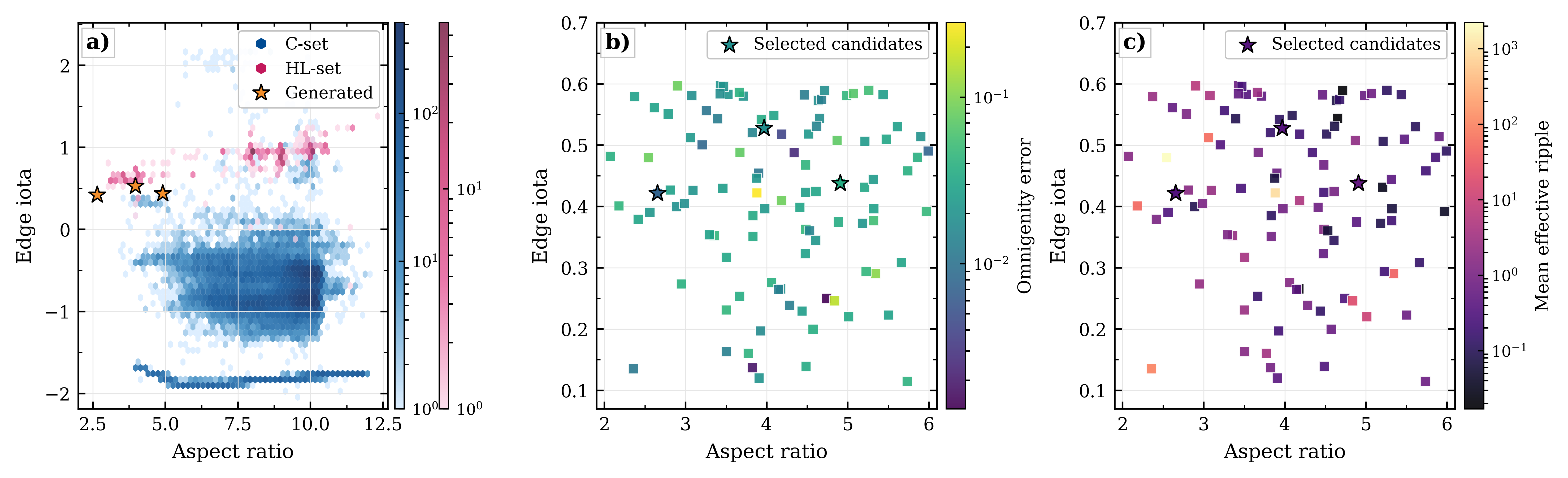}
    \caption{Two-dimensional parameter-scan results for compact low-rotational-transform configurations. (a) Data distribution of the C set and original HL set in the \(A\)-\(\iota\) plane; stars indicate the three representative generated configurations further analyzed in Fig.~\ref{fig:representative_configs}. (b) Heat map of omnigenity error for generated candidate configurations. (c) Heat map of average effective ripple for generated candidate configurations.}
    \label{fig:scan_results}
\end{figure}

As shown in Fig.~\ref{fig:scan_results}, the C set and original HL
set distributions in the \(A\)-\(\iota\) plane indicate that the low-aspect-ratio
and low-rotational-transform region is poorly covered by the original data.
After full-model low-learning-rate fine-tuning with source-domain replay, the model can
generate DESC-verified converged candidate configurations in this sparse region, and
their realized magnetic-geometric indicators remain consistent with the target conditions.
Figure~\ref{fig:scan_results}c further shows the effective-ripple distribution of generated samples in
this parameter plane, providing a transport-related supplementary quality assessment. Some
ultra-compact generated configurations simultaneously show low effective ripple, suggesting that
the adapted model not only expands candidate coverage in the
low-\(A\) region, but also provides initial samples with promising transport
potential for subsequent high-fidelity QI optimization.

\begin{figure}[t]
    \centering
    \includegraphics[width=0.95\textwidth]{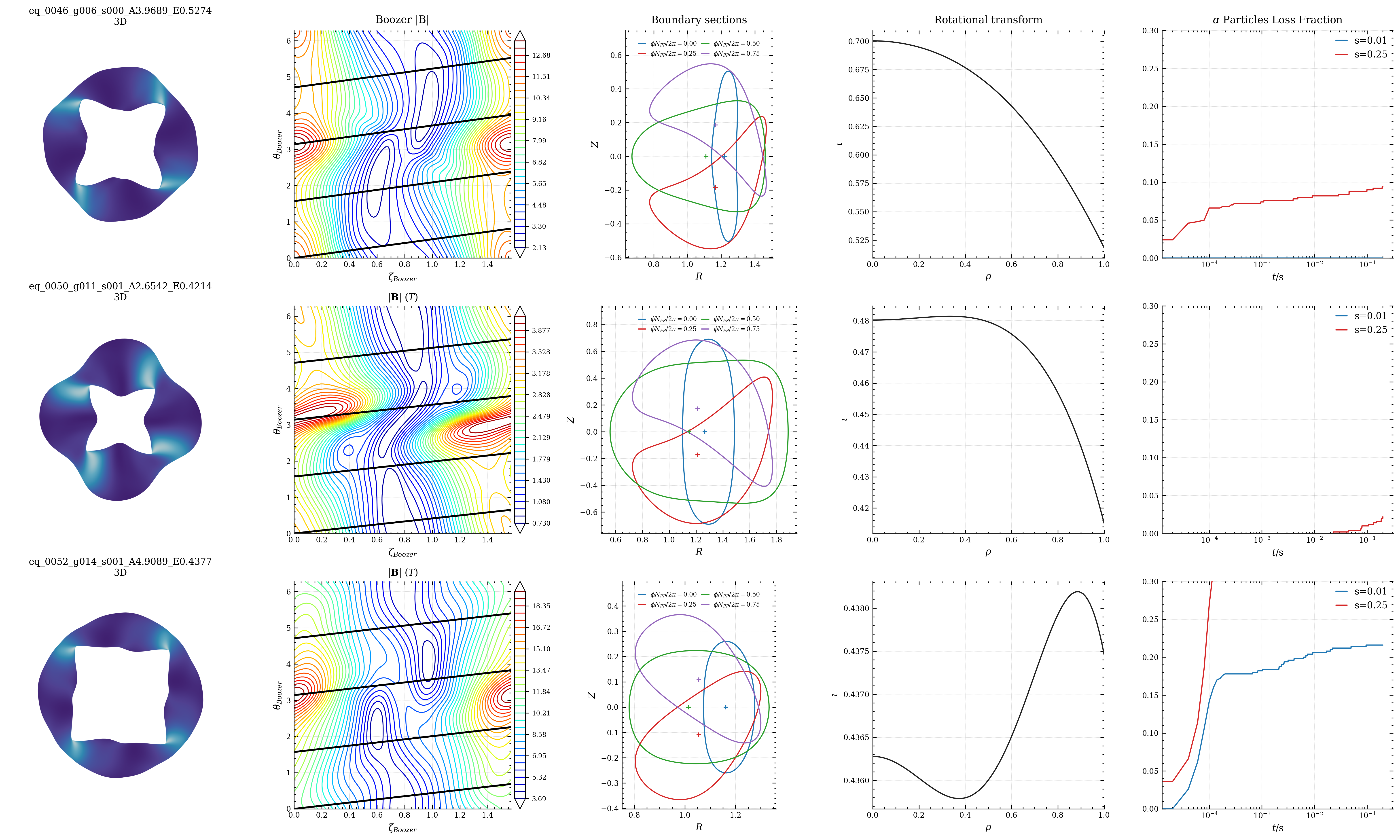}
    \caption{MHD equilibrium and magnetic-field-structure analysis of three high-quality generated configurations selected using omnigenity error, effective ripple, and Boozer-plot properties. From left to right, the columns show the three-dimensional equilibrium surface, contours of \(\lvert B\rvert\) in Boozer coordinates, boundary cross-sections at four toroidal angles, the rotational-transform profile, and the cumulative \(\alpha\)-particle loss fraction at \(s=0.01\) and \(s=0.25\), respectively.}
    \label{fig:representative_configs}
\end{figure}

We further select three high-quality representative configurations from the generated
samples for detailed analysis; these are marked by stars in
Fig.~\ref{fig:scan_results}a. The selection criteria combine DESC fixed-boundary equilibrium convergence, small
target magnetic-geometric errors, ultra-compactness (\(A<5\)), omnigenity error, effective ripple,
and magnetic-field-strength contour structure in Boozer plots. The three samples
shown in Fig.~\ref{fig:representative_configs} perform well under these diagnostics and
can serve as initial candidates for subsequent high-fidelity QI optimization
and coil-feasibility assessment.

Table~\ref{tab:representative_metrics} gives the quantitative comparison of these
three configurations. All three DESC solutions have normalized force-balance
residuals of order \(10^{-2}\), while their QI residual, effective ripple, and
fast-particle confinement differ appreciably. The normalized current-density
metric is reported separately from the force-balance residual because it
characterizes the magnitude of the current-density objective, rather than the
MHD force imbalance itself. In particular, the existing SIMPLE calculations
give zero core alpha-particle loss through \(t=0.1~\mathrm{s}\) for
\(\mathrm{eq\_0046}\) and \(\mathrm{eq\_0050}\), whereas
\(\mathrm{eq\_0052}\) has a loss fraction of 0.216.

\begin{table}[H]
    \centering
    \caption{DESC-based magnetic-geometric, equilibrium, and confinement metrics for the three representative generated configurations. \(\overline{\epsilon}_{\mathrm{QI}}\) is the mean absolute OOPS residual, \(\mathcal{R}_F=\langle\lvert\mathbf{J}\times\mathbf{B}-\nabla p\rvert\rangle_V/\langle\lvert\nabla(B^2)\rvert/(2\mu_0)\rangle_V\) is the normalized force-balance residual, and \(\overline{J}_{\mathrm{norm}}\) is the normalized mean absolute DESC CurrentDensity objective. The alpha-particle loss fraction \(f_\alpha\) is evaluated from the existing SIMPLE trajectories at \(s=0.01\) and \(t=0.1~\mathrm{s}\); \(\langle\epsilon_{\mathrm{eff}}\rangle_r\) is the arithmetic mean of finite NEO values over the sampled radial surfaces.}
    \label{tab:representative_metrics}
    \vspace{2pt}
    {\renewcommand{\arraystretch}{1.18}%
    \resizebox{\textwidth}{!}{%
    \begin{tabular}{lcccccccccc}
        \toprule
        Configuration & \(A\) & \(\iota_{\mathrm{edge}}\) & \(\kappa_{\max}\) & \(\overline{\delta}\) & \makecell{\(\overline{\epsilon}_{\mathrm{QI}}\)\\(OOPS)} & \(\mathcal{R}_F\) & \(\overline{J}_{\mathrm{norm}}\) & \(N_{\mathrm{fp}}\) & \makecell{\(f_\alpha\)\\\(s=0.01,\ t=0.1~\mathrm{s}\)} & \(\langle\epsilon_{\mathrm{eff}}\rangle_r\) \\
        \midrule
        \(\mathrm{eq\_0046}\) & 3.969 & 0.518 & 6.68 & -0.236 & \(1.80\times10^{-2}\) & \(1.39\times10^{-2}\) & \(2.24\times10^{-3}\) & 4 & 0.000 & 0.348 \\
        \(\mathrm{eq\_0050}\) & 2.654 & 0.415 & 3.38 & -0.011 & \(8.59\times10^{-3}\) & \(1.39\times10^{-2}\) & \(1.64\times10^{-3}\) & 4 & 0.000 & 0.313 \\
        \(\mathrm{eq\_0052}\) & 4.909 & 0.437 & 2.58 & -0.235 & \(3.26\times10^{-2}\) & \(1.03\times10^{-2}\) & \(9.34\times10^{-4}\) & 4 & 0.216 & 0.675 \\
        \bottomrule
    \end{tabular}%
    }
    }
\end{table}

Among these candidates, \(\mathrm{eq\_0046}\) and \(\mathrm{eq\_0050}\) are
particularly noteworthy because both retain zero core \(\alpha\)-particle loss
at \(s=0.01\) through \(t=0.1~\mathrm{s}\) in the existing SIMPLE diagnostic.
\(\mathrm{eq\_0050}\) has the lowest aspect ratio of the representative set and
the smallest OOPS residual and radial-mean effective ripple, whereas the
ultra-compact vacuum configuration \(\mathrm{eq\_0046}\) is selected as the
initial condition for the downstream finite-\(\beta\) QI optimization. Their
converged vacuum equilibria and favorable core fast-particle confinement make
both configurations promising candidates for subsequent high-fidelity optimization
and coil-feasibility assessment.

\subsection{Downstream Vacuum QI Optimization and Finite-\texorpdfstring{\(\beta\)}{beta} Assessment}

We next assess whether a diffusion-generated candidate can provide a useful
entry point to a higher-fidelity optimization workflow. The generated vacuum
configuration \(\mathrm{eq\_0046}\) is initially ultra-compact, with
\(A=3.969\). The subsequent vacuum QI boundary optimization reduces the
objective over 200 iterations while increasing the aspect ratio to \(A=10.374\). This loss of
compactness occurs during the vacuum boundary optimization, where aspect ratio
is unconstrained, rather than during the later finite-\(\beta\) calculation.
The finite-\(\beta\) calculation retains this optimized boundary and reaches a
volume-averaged beta of \(\langle\beta\rangle=5.06\%\).

\begin{figure}[t]
    \centering
    \includegraphics[width=0.95\textwidth]{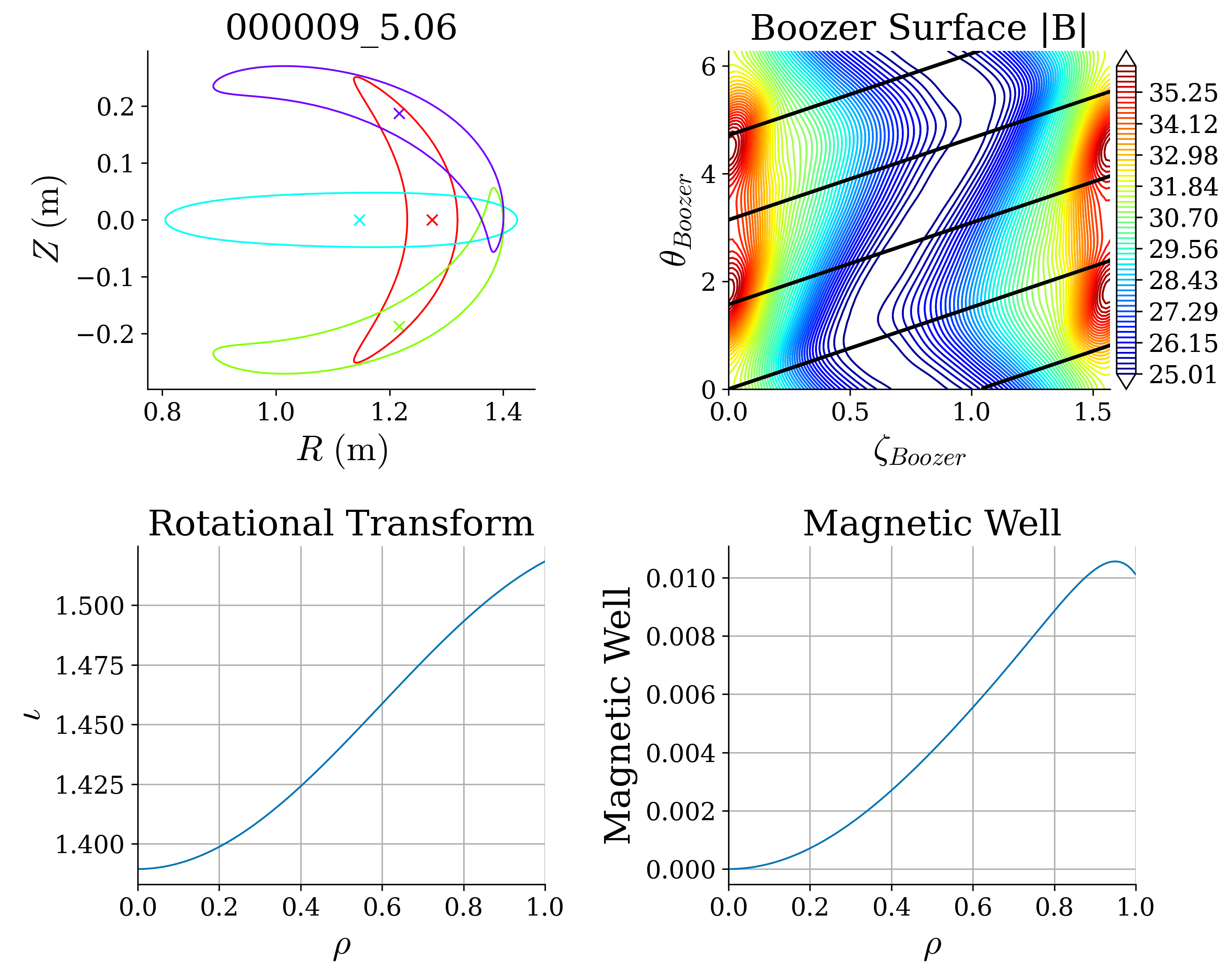}
    \caption{Equilibrium analysis of the fixed-boundary finite-\(\beta\) configuration with \(\langle\beta\rangle=5.06\%\). The panels show boundary cross-sections at four toroidal angles, \(\lvert B\rvert\) contours in Boozer coordinates with representative field lines, the rotational-transform profile, and the magnetic-well profile.}
    \label{fig:finite_beta_equilibrium}
\end{figure}

Figure~\ref{fig:finite_beta_equilibrium} summarizes the equilibrium properties
of the finite-\(\beta\) configuration. The rotational transform increases from
\(\iota=1.389\) on axis to \(\iota=1.518\) at the edge, and the magnetic-well
profile is positive away from the axis, reaching 0.0106. The boundary
cross-sections and Boozer-coordinate \(\lvert B\rvert\) contours remain smooth
at this pressure. These results demonstrate that the QI-optimized boundary
obtained from the generated candidate supports a converged finite-\(\beta\)
equilibrium, although it is no longer ultra-compact.

\begin{figure}[p]
    \centering
    \includegraphics[width=0.95\textwidth]{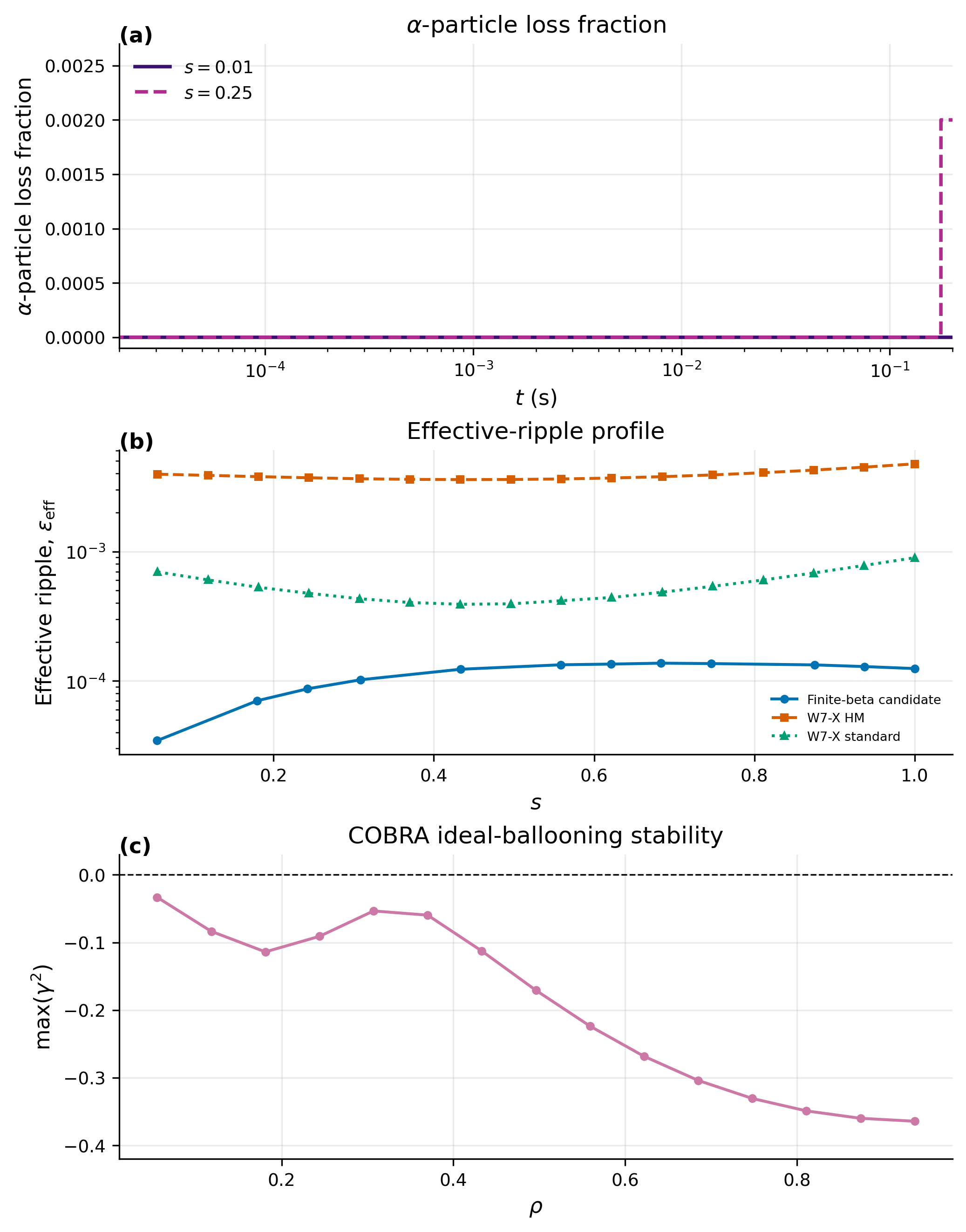}
    \caption{Confinement and ideal-ballooning diagnostics for the finite-\(\beta\) configuration. (a) Cumulative \(\alpha\)-particle loss fraction from SIMPLE tracking of 500 particles initialized at \(s=0.01\) and \(s=0.25\). (b) NEO effective-ripple profile; the W7-X High-Mirror (HM) and standard reference curves are included for comparison. (c) Maximum COBRA ideal-ballooning \(\gamma^2\) over the sampled poloidal and field-line labels as a function of normalized radius; negative values indicate stability within the sampled grid.}
    \label{fig:finite_beta_diagnostics}
\end{figure}

The supplementary confinement and stability diagnostics are shown in
Fig.~\ref{fig:finite_beta_diagnostics}. SIMPLE tracking of 500 \(\alpha\)
particles to \(t=0.2~\mathrm{s}\) gives zero loss from \(s=0.01\) and a loss
fraction of 0.002 from \(s=0.25\). The NEO effective ripple remains between
approximately \(3.5\times10^{-5}\) and \(1.4\times10^{-4}\) over the sampled
surfaces and lies below both W7-X reference curves plotted with the same
diagnostic workflow. In the COBRA calculation, the maximum ideal-ballooning
\(\gamma^2\) is negative at every sampled interior radius, with a least-negative
value of \(-0.033\). This indicates the absence of an ideal-ballooning
instability on the sampled radial, poloidal, and field-line grids; it is not a
global MHD-stability proof.

This case study establishes a limited but direct downstream validation: the
generative model supplies an ultra-compact vacuum seed from which a QI-optimized
boundary supporting a favorable finite-\(\beta\) equilibrium can be obtained.
It does not demonstrate a finite-\(\beta\), ultra-compact final configuration.
Achieving both objectives requires an explicit compactness term or constraint
in the vacuum QI boundary optimization.

\section{Discussion}

The results show that the VAE latent-diffusion framework is well suited to
high-dimensional, strongly coupled three-dimensional boundary generation for QI
stellarators. Stellarator boundary Fourier coefficients are not arbitrary independent
variables; they are constrained by boundary smoothness, field-period symmetry, nested
flux surfaces, and MHD equilibrium solvability. Their effective degrees of freedom are
therefore closer to a nonlinear low-dimensional manifold, motivating representations
learned directly from databases of usable equilibria.

In the reconstruction comparison used here, PCA seeks a low-dimensional linear subspace
in coefficient space, with retained information determined primarily by sample variance.
It cannot fully capture nonlinear coupling among Fourier modes, and directions with large
coefficient-space variance need not correspond to the most important degrees of freedom
in physical boundary geometry. In contrast, the VAE uses a nonlinear encoder--decoder
structure and an \(R/Z\) physical-space reconstruction loss, so latent learning is
directly constrained by boundary-geometry error. This accounts for the better VAE
reconstruction accuracy and boundary visualization in the present benchmark and provides
a suitable latent space for conditional diffusion. This conclusion is limited to the
present reconstruction benchmark; reconstruction accuracy alone does not determine
performance in downstream optimization.

The core value of the conditional diffusion model is to
transform high-dimensional random search in stellarator design into conditional candidate
generation in a low-dimensional latent space. For the same target
aspect ratio, triangularity, elongation, and edge rotational transform, there may
exist multiple three-dimensional boundaries that satisfy the conditions but differ
substantially in physical quality. Thus, the proposed ``conditional generation--high-fidelity equilibrium
verification--candidate screening'' workflow is better viewed as a front-end component
for DESC/VMEC, subsequent QI optimization, and coil optimization, rather than
as a replacement for those physics solvers. Its role is
to narrow the search space that must be processed by
expensive solvers and to concentrate computational resources on more promising
candidate initial conditions. The generated candidates can then be passed to DESC/VMEC refinement as part of a broader design workflow.

The domain-adaptation results show that boundary-generation priors learned from
large-scale ordinary-QI data can be transferred to the sample-sparse
ultra-compact region. These configuration classes exhibit clear distributional differences,
especially in low-\(A\) and low-\(\iota\) regions, where a directly applied
pretrained model tends to suffer from conditional mismatch or validation
failure. Low-learning-rate fine-tuning corrects the local latent distribution and conditional
mapping in the new domain, while source-domain replay mitigates catastrophic
forgetting and preserves more general QI geometric knowledge from the
source domain. Methodologically, this indicates that domain adaptation is not
simply additional training, but a recalibration of the conditional generative
distribution in a data-sparse target region.

Physically, ultra-compact designs require stronger three-dimensional shaping and
involve more complex rotational-transform responses and mode coupling. In this
region, simultaneously obtaining a converged MHD equilibrium, target magnetic-geometric consistency,
low effective ripple, and favorable particle confinement is difficult for
traditional design and optimization methods. The representative configurations identified in
this work indicate that the latent generative model captures part
of the effective geometric structure of compact QI boundaries and
can propose candidates of further optimization value in low-aspect-ratio regimes.
The downstream calculation based on \(\mathrm{eq\_0046}\) provides a more direct
test of this role. The generated configuration begins as an ultra-compact vacuum
candidate with \(A=3.969\), but the unconstrained vacuum QI boundary optimization
increases the aspect ratio to 10.374. This increase occurs before pressure is
introduced: the subsequent finite-\(\beta\) calculation fixes that optimized
boundary and obtains \(\langle\beta\rangle=5.06\%\). The result therefore
validates the generated configuration as an entry point to a higher-fidelity
optimization workflow, rather than as a demonstration of a finite-\(\beta\),
ultra-compact final state.

At \(\langle\beta\rangle=5.06\%\), the fixed-boundary equilibrium retains a
positive magnetic well away from the axis and a monotonic rotational-transform
profile. The physical diagnostics are also favorable within their stated
scope: SIMPLE tracking gives zero loss for particles initialized at
\(s=0.01\) through \(0.2~\mathrm{s}\), NEO yields
\(\epsilon_{\mathrm{eff}}\) of order \(10^{-4}\) or below over the sampled
surfaces, and COBRA gives no positive ideal-ballooning \(\gamma^2\) on the
sampled grid. These checks do not establish global MHD stability, reactor-scale
particle confinement, or coil feasibility. They do, however, show that the
generated initial condition can lead to a QI-optimized boundary compatible with
a converged finite-\(\beta\) equilibrium and several favorable first-pass
physics diagnostics.

The case also identifies a clear limitation of the current workflow. Since
compactness is absent from the vacuum QI boundary-optimization objective, the
downstream optimization improves the targeted physics objectives at the expense
of aspect ratio. Future work should therefore incorporate an aspect-ratio target
or bound together with finite-\(\beta\), transport, stability, and coil
constraints, so that the optimization searches directly for configurations that
retain compactness while satisfying the relevant physical requirements.

Several limitations remain. First, the original HL set is small,
and the augmented HL set remains localized near existing high-fidelity
configurations, leaving more extreme compact regimes insufficiently covered. Second, the
current condition vector includes only a limited set of magnetic-geometric
indicators and does not directly incorporate particle loss, effective ripple,
MHD stability, magnetic islands, or coil feasibility. Future work should
include effective ripple, fast-particle constraints, and quantitative Boozer-plot structure metrics
that are more suitable for cross-configuration comparison, and should couple
VAE-Diffusion with high-fidelity equilibrium optimization and coil optimization to form
a more complete closed-loop design workflow for ultra-compact QI stellarators.

\section{Conclusion and Future Work}

We proposed and validated a domain-adaptive latent diffusion framework for
ultra-compact four-field-period QI stellarator configuration generation. The framework first uses
a VAE with a physical-space \(R/Z\) reconstruction constraint to learn
a low-dimensional nonlinear latent space for three-dimensional boundary Fourier coefficients,
and then trains a conditional diffusion model in this latent
space to generate three-dimensional plasma-boundary candidates from target magnetic-geometric indicators.
The framework learns a conditional distribution and addresses small-sample distribution
shift in the ultra-compact QI region. The model can therefore serve as both a rapid scanner of
ultra-compact design space and an initial-condition generator for high-fidelity optimization.

The main conclusions are as follows. First, the VAE compresses
and reconstructs stellarator boundaries more effectively than linear methods such
as PCA. At the selected 43-dimensional working point, the model
maintains good Fourier-coefficient reconstruction accuracy and physical boundary-geometry fidelity, indicating
that stellarator boundary data can be approximated by a low-dimensional
nonlinear manifold. Second, the conditional diffusion model trained in this
latent space is stable, with a test-set noise-prediction MSE of
approximately 0.1004. Generated samples under unseen target conditions and subsequent
DESC verification yield a combined mean percentage error of about
4.2\% across the four magnetic-geometric indicators, showing good conditional controllability.
Third, domain adaptation based on the augmented HL set substantially
mitigates distribution shift in the ultra-compact target domain. Full-model low-learning-rate
fine-tuning with source-domain replay gives the best performance, reducing compact
test loss from 3.2596 to 0.4147, a relative improvement of
about 87.28\%. Fourth, after adaptation, the model generates converged ultra-compact
QI candidates in sparse low-aspect-ratio and low-rotational-transform regions. Diagnostics based
on effective ripple, Boozer-plot properties, and particle loss identify representative
samples with distinct confinement trade-offs. In particular, \(\mathrm{eq\_0046}\)
and \(\mathrm{eq\_0050}\) retain zero core \(\alpha\)-particle loss through
\(t=0.1~\mathrm{s}\) in the present diagnostic, while \(\mathrm{eq\_0050}\)
combines the lowest aspect ratio of the representative set with the smallest
OOPS residual and a radial-mean effective ripple of 0.313. A downstream
vacuum QI boundary optimization initialized from \(\mathrm{eq\_0046}\) produces
a boundary with \(A=10.374\), which subsequently supports a fixed-boundary
finite-\(\beta\) equilibrium at \(\langle\beta\rangle=5.06\%\). At this beta,
the SIMPLE, NEO, and sampled COBRA diagnostics respectively show zero core
\(\alpha\)-particle loss through \(0.2~\mathrm{s}\), effective ripple of order
\(10^{-4}\) or below, and no positive ideal-ballooning \(\gamma^2\) on the
sampled grid. This demonstrates the utility of the generated vacuum initial
condition while also revealing that the loss of compactness occurs in the
unconstrained downstream vacuum QI boundary optimization, not in the
fixed-boundary finite-\(\beta\) calculation.

These results indicate that the proposed VAE-Diffusion framework has clear
innovation and practical value. Its key contribution is not merely
the use of a diffusion model, but the integration of
latent-space conditional generation, domain adaptation, and high-fidelity physics verification into
an executable candidate-exploration workflow for the sample-sparse ultra-compact QI region.
The framework demonstrates a complete path from design-space scanning to
high-quality initial-condition proposal. It can provide focused candidate starting points
for high-fidelity QI optimization, particle-confinement assessment, and coil-feasibility optimization, and
can also support sample selection in active-learning design loops.

Future work will improve the physical completeness and engineering usability
of the model. Effective ripple, fast-particle constraints, MHD stability, Boozer-plot
structure, and coil feasibility can be incorporated into the condition
vector or sampling guidance. Active learning can be used to automatically enrich
high-value training samples in the ultra-compact region, with explicit weighting to
prevent the enlarged source domain from overwhelming the target-domain prior. The
generative model can also be coupled with bounded data-informed Bayesian optimization,
DESC/VMEC refinement, and coil-optimization tools to build an end-to-end closed-loop
workflow from target performance to boundary and then to coil design. This comparison
should evaluate not only reconstruction and conditional-target errors, but also the
usable fraction of generated or searched boundaries, equilibrium-convergence rate, and
the number of high-fidelity objective evaluations required to reach a prescribed
physics threshold. In particular, aspect ratio should be retained as an explicit
downstream constraint when compact finite-\(\beta\) configurations are sought. As
high-fidelity QI datasets continue to expand, the proposed method can further develop
into a rapid, scalable, and physically consistent data-driven platform for advanced
stellarator conceptual design.

\section*{Acknowledgements}

This work was supported by the National Natural Science
Foundation of China under Grant No. 12405274, the Anhui Provincial Key Research and Development Project under
Grant No. 2023a05020008, and the Strategic Priority Research Program of the Chinese Academy of
Sciences with Grant No. XDB0790302. The authors would like to thank colleagues and collaborators for
helpful discussions on stellarator optimization, quasi-isodynamic configuration design, and
machine-learning-based inverse design methods. The authors also acknowledge the developers
and maintainers of the open-source computational tools and datasets used
in this work, including DESC and ConStellaration.

\appendix
\section*{Appendix}
\addcontentsline{toc}{section}{Appendix}

\section{Omnigenity-Error Metric}\label{app:omni_error}

The omnigenity error used in this work follows the Optimized Parallels
(OOPS) formulation of Liu \emph{et al.}\ \cite{liu2025omnigenity}. In OOPS
coordinates, the magnetic-field strength of an ideally omnigenous field depends
only on the coordinate \(\eta\), denoted by \(B_{\mathrm{omni}}(\eta)\), whereas
the field strength of an actual equilibrium, \(B_{\mathrm{eq}}\), generally depends
on the Boozer angles \((\theta_B,\zeta_B)\). For the four-field-period QI
configurations considered here, we use helicity \(\bm{h}=(0,N_{\mathrm{fp}})\).
The OOPS mapping from \((\eta,\alpha)\) to Boozer coordinates is
\begin{align}
\theta_B &= N_{\mathrm{fp}}\alpha+\frac{\iota}{N_{\mathrm{fp}}}h, \\
\zeta_B &= \frac{h}{N_{\mathrm{fp}}},
\end{align}
where \(\iota\) is the boundary rotational transform and \(h\) is the
omnigenity symmetry angle. Reconstructing \(B_{\mathrm{eq}}\) from its Boozer
Fourier spectrum \(\lvert B\rvert_{mn}^{B}\), the pointwise omnigenity residual is
defined as
\begin{equation}
\epsilon_{\mathrm{omni}}(\eta,\alpha)
= B_{\mathrm{eq}}\!\left(\theta_B(\eta,\alpha),\zeta_B(\eta,\alpha)\right)
- B_{\mathrm{omni}}(\eta).
\end{equation}
For an ideal QI configuration, \(B_{\mathrm{eq}}\) depends only on \(\eta\),
and this residual vanishes. The reported mean omnigenity error is the
mean absolute residual over the \((\eta,\alpha)\) grid:
\begin{equation}
\overline{\epsilon}_{\mathrm{omni}}
= \frac{1}{N_{\eta}N_{\alpha}}
\sum_{i=1}^{N_{\eta}}\sum_{j=1}^{N_{\alpha}}
\left\lvert\epsilon_{\mathrm{omni}}(\eta_i,\alpha_j)\right\rvert,
\end{equation}
where \(N_{\eta}\) and \(N_{\alpha}\) are the numbers of grid points in the
two OOPS coordinates.

\section{Equilibrium Perturbation Expansion for Data Augmentation}

We use the perturbation and continuation methods in DESC to
locally augment the original HL set \cite{conlin2023desc2}. Let the original
equilibrium solution be \(x_0\), the control parameters be \(p_0\), and
the equilibrium equation be \(F(x,p)=0\). After applying a small perturbation
\(\Delta p\) near \((x_0,p_0)\), the response is written as
\begin{equation}
\Delta x=\Delta x^{(1)}+\Delta x^{(2)}.
\end{equation}
A second-order Taylor expansion of \(F(x_0+\Delta x,p_0+\Delta p)=0\) gives
\begin{equation}
F_x\Delta x+F_p\Delta p+
\frac{1}{2}\left(F_{xx}[\Delta x,\Delta x]
+2F_{xp}[\Delta x,\Delta p]
+F_{pp}[\Delta p,\Delta p]\right)\approx 0 .
\end{equation}
Balancing terms at different orders yields the first-order response
\begin{equation}
\Delta x^{(1)}=-F_x^{-1}F_p\Delta p,
\end{equation}
and the second-order response
\begin{equation}
\Delta x^{(2)}
=-\frac{1}{2}F_x^{-1}
\left(F_{xx}[\Delta x^{(1)},\Delta x^{(1)}]
+2F_{xp}[\Delta x^{(1)},\Delta p]
+F_{pp}[\Delta p,\Delta p]\right).
\end{equation}
In the augmentation procedure, each original HL configuration generates three
perturbed samples. Boundary Fourier coefficients \(R_{b,mn}\) and \(Z_{b,mn}\) are perturbed
by approximately 10\% relative Gaussian noise, with high-frequency modes damped
to preserve boundary smoothness. Pressure profiles, rotational-transform profiles, and total
toroidal flux are perturbed by approximately 5\%, 3\%, and 2\%,
respectively. The perturbation order is set to 2 and the
trust-region ratio to 0.1, followed by a DESC solve to
restore force balance.

\section{Physical-Space Reconstruction Error for the VAE}

To ensure that the VAE latent variables preserve not only
Fourier-coefficient statistics but also true boundary geometry, we add the
\(R/Z\) physical-space collocation reconstruction loss during training. The collocation grid
is generated from boundary angular coordinates: the poloidal angle \(\theta\)
is uniformly sampled in \([0,2\pi)\), and the toroidal angle \(\phi\)
is uniformly sampled over one field period \([0,2\pi/N_{\mathrm{fp}})\), where \(N_{\mathrm{fp}}=4\)
in this work. A \(32\times32\) grid is used during training,
and a \(64\times64\) grid is used during evaluation.

The Fourier basis matrices \(\Psi_R(\theta,\phi)\) and \(\Psi_Z(\theta,\phi)\) are constructed from
the DESC boundary Fourier-mode indices. Given boundary coefficients, the radial
and vertical coordinates on the collocation grid are
\begin{equation}
R(\theta,\phi)=\bm{x}_R\Psi_R^\top(\theta,\phi),\qquad
Z(\theta,\phi)=\bm{x}_Z\Psi_Z^\top(\theta,\phi).
\end{equation}
Decoded Fourier coefficients give \(\hat R\) and \(\hat Z\) in
the same way, from which \(\mathcal{L}_{RZ}\) is computed. This term
directly penalizes geometric reconstruction error in physical boundary space while
also reducing coefficient-space error.

\section{Distribution-Consistency Validation of the Conditional Diffusion Model}

\begin{figure}[t]
    \centering
    \includegraphics[width=0.90\textwidth]{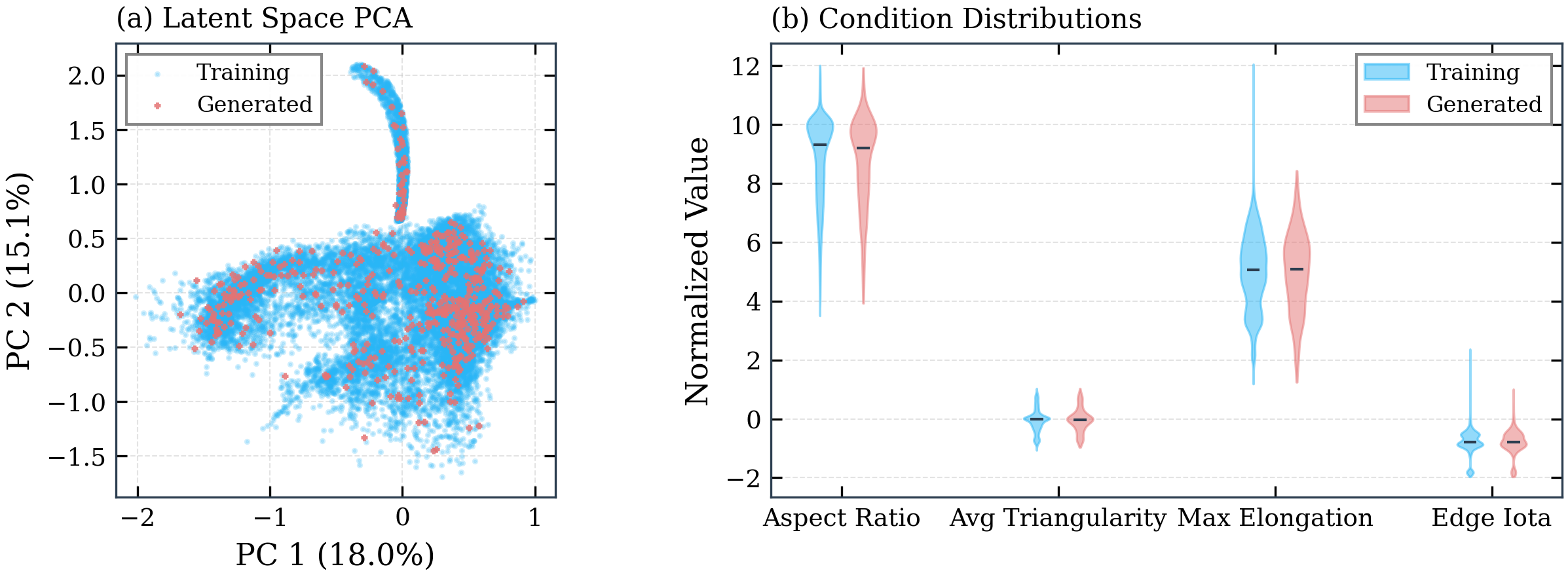}
    \includegraphics[width=0.90\textwidth]{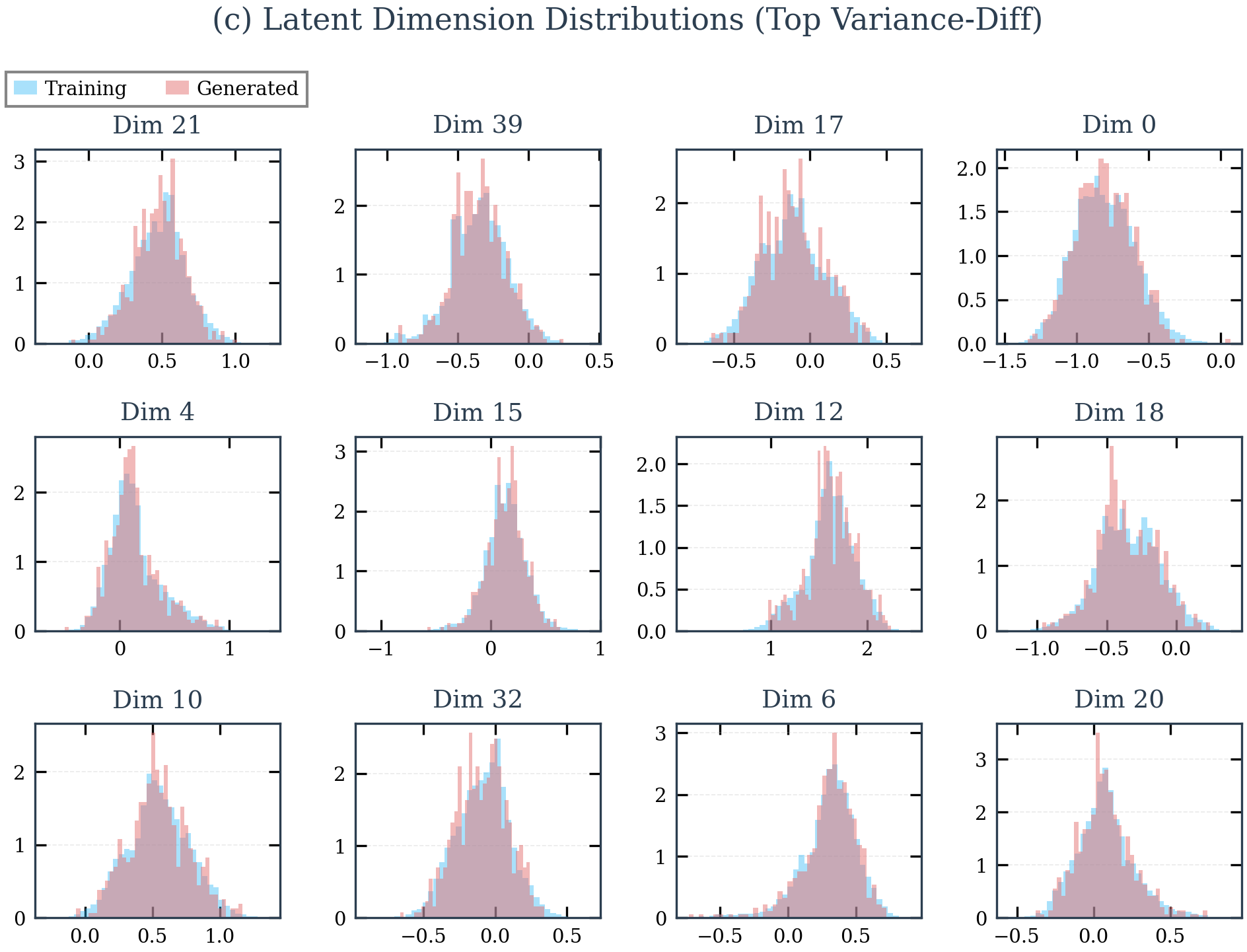}
    \caption{Distribution-consistency validation of CDDPM-generated samples.}
    \label{fig:appendix_cddpm}
\end{figure}

We further validate the conditional diffusion model from the perspective
of distribution consistency under the selected hyperparameter configuration (hidden dimension
512, 10 layers, learning rate \(1\times10^{-3}\)), as shown in Fig.~\ref{fig:appendix_cddpm}.
The validation-set noise-prediction MSE is approximately 0.096, the test-set MSE
is approximately 0.100, and the best performance is reached at
about epoch 472, indicating good convergence and generalization in the
43-dimensional latent space. Figure~\ref{fig:appendix_cddpm}a compares the training and generated samples
projected onto VAE latent-space PCA coordinates; generated samples overlap strongly
with training samples and show no obvious mode collapse or
distributional shift. Figure~\ref{fig:appendix_cddpm}b quantitatively compares distributions of the four target
magnetic-geometric indicators: mean deviations are approximately 1.1\% for aspect ratio,
0.2\% for maximum elongation, and 0.7\% for edge rotational transform,
and all relative standard-deviation deviations are within 5\%. Figure~\ref{fig:appendix_cddpm}c compares
the 12 latent dimensions with the largest variance differences. The
generated and training distributions generally agree in peak locations and
spread. These results show that the CDDPM is not only
stable in noise prediction, but also preserves latent-manifold and conditional-distribution
statistics, providing reliable candidates for VAE decoding and DESC equilibrium
verification.

\bibliographystyle{unsrtnat}
\bibliography{references}

\end{document}